# Ferroelectric nematic – isotropic liquid critical end point

Jadwiga Szydlowska[1], Pawel Majewski[1], Mojca Čepič, [2,3] Nataša Vaupotič, [2,4] Paulina Rybak,[1] Corrie T Imrie[5], Rebecca Walker[5], Ewan Cruickshank[5], John MD Storey[5], Damian Pociecha,[1] and Ewa Gorecka[1*]

[1] Faculty of Chemistry, University of Warsaw, Żwirki i Wigury 101, 02-089 Warsaw, Poland
[2] Jozef Stefan Institute, Jamova 39, 1000 Ljubljana, Slovenia
[3] Department of Physics and Technical Studies, Faculty of Education, University of Ljubljana, Kardeljeva ploščad 16, 1000 Ljubljana, Slovenia
[4] Department of Physics, Faculty of Natural Sciences and Mathematics, University of Maribor, Koroška 160, 2000 Maribor, Slovenia
[5] Department of Chemistry, University of Aberdeen, Old Aberdeen AB24 3UE, U.K.

**Abstract:** A critical end point above which an isotropic phase continuously evolves into a polar (ferroelectric) nematic phase with an increasing electric field is found in a ferroelectric nematic liquid crystalline material. The critical end point is approximately 30 K above the zero-field transition temperature from the isotropic to nematic phase and at an electric field of the order of 10 V/μm. Such systems are interesting from the application point of view because a strong birefringence can be induced in a broad temperature range in an optically isotropic phase.

The discovery of ferroelectricity in the least ordered liquid crystalline phase, the nematic phase, [1-7] immediately attracted considerable attention, because ferroelectricity and fluidity were thought of as being mutually exclusive properties as the long-range ordering of dipole moments could easily be suppressed by thermal fluctuations in the liquid state. A polar liquid is an interesting subject of fundamental studies but ferroelectricity and flexibility are also attractive from the point of view of future applications. In most of the systems studied so far, the ferroelectric nematic phase ($N_F$) appears below an apolar nematic phase. The transition between these phases is weakly first or second order; a polar structure gradually develops in the higher temperature nematic phase as evidenced by a strong dielectric response that critically slows down on approaching the polar phase [8]. To date, the $N_F$ phase has been reported for a limited number of materials, but it seems that all of these bear some general features – they have a strong longitudinal dipole moment and bulky lateral substituents. While it is obvious why a strong dipole moment is necessary, the role of lateral substituents is still under debate. On one hand, they might introduce a wedge-like molecular shape which results in a lower splay elastic constant and an increase of the splay flexoelectric coefficient in such materials. On the other hand, lateral substituents may prevent antiparallel correlations of molecular pairs in the transverse direction, which is necessary to obtain a long-range parallel orientation of dipole moments. Recently, the first examples of systems in which the $N_F$ phase appears directly below the isotropic liquid phase were reported [5, 9-11]. Systematic studies showed that the Iso-N phase transition temperature decreases more quickly than the $N_F$-N phase transition temperature when the size of lateral substituents increases, and as a result, an Iso-$N_F$ transition was found for some materials or a sequence Iso-N-$N_F$ in which the N phase exists in a very short temperature range [11]. In both types of systems, the $N_F$ phase develops by a first-order transition. The availability of ferroelectric liquid crystal compounds showing a first-order N-$N_F$ or Iso-$N_F$ phase transition enables an investigation of an electric-field-induced critical point, which we report in this study. We show that the first-order transition that is characterized by a discontinuity of the order parameter – electric polarization - remains first-order at moderate electric field strengths and vanishes at higher fields, implying the existence of a critical end point in the temperature-electric field phase diagram.

Several compounds showing either a short temperature range of a paraelectric nematic phase preceding a ferroelectric $N_F$ or a direct $N_F$-Iso transition were investigated. They all show either a strongly first-order transition between both nematic phases evidenced by a large jump in birefringence at this transition or the first-order direct N_F-Iso phase transition. It should be mentioned that in all the materials studied so far, the N_F phase is monotropic but the accessible temperature range over which the properties of the system could be studied differ significantly for each compound because some of the studied compounds readily crystallized. Therefore, although the overall behavior of the systems studied was similar, for detailed studies we chose a chiral material I-4*, see Fig. S1 in Supplementary Material (SM) [12] with a short paraelectric nematic (cholesteric) phase (~1 K) and an N_F phase that could easily be supercooled to room temperature. Importantly, this material showed the lowest tendency towards cold crystallization when heated above the clearing point. To exclude a possible influence of the molecular chirality on the polar properties of the studied system we have compared the phase transition temperatures of a pure enantiomer and racemic mixture. The polar nematic phase exists in both and the transition temperatures differ by less than 0.2 K (see SM).

For electro-optic studies, the material was filled into a glass cell with transparent ITO electrodes covered with an aligning polymer layer. In the N_F phase under the application of an a.c. voltage, a single repolarization current peak per half of a cycle was registered, characteristic of the ferroelectric phase, but as temperature increased and an apolar nematic or isotropic liquid phase was entered, the current peak doubled and the voltage of the onset of the current peaks increased (Fig. 1 and Figs. S2 and S3 in SM). The polarization calculated from a single current peak in the N_F phase and from a double current peak just above the N_F-Iso phase was very similar and diminished on further heating. Optical measurements show that the electric field induces a birefringence in the isotropic liquid phase, which increases slightly with the applied voltage (low field part of isotherms). At a field corresponding to the current peak onset (threshold field, $E_{th}$) a step-like increase of optical birefringence is observed, marked by an increase of the light transmission through the sample between crossed polarizers (Fig. 1). Thus, the electric field induces a polar order in an isotropic liquid, which is coupled to the orientational order of the molecules. By decreasing the electric field, the birefringence changes are reversed with a small hysteresis.

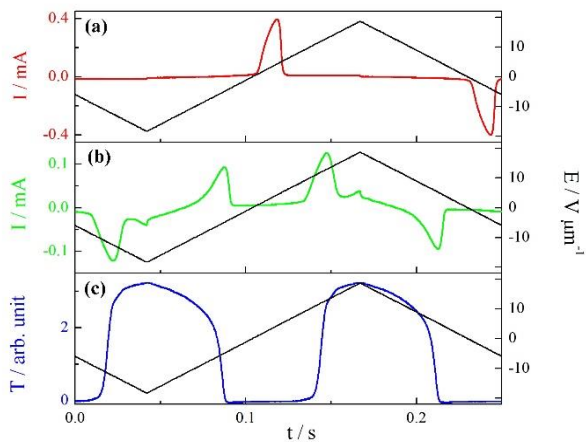

Figure 1. a) A single repolarization current peak recorded as a function of time ($t$) in the N_F phase and b) a double current peak recorded in the isotropic liquid, under application of a triangular voltage (black line, right scales). c) The light transmission ($T$) through a sample placed between crossed polarizers, taken in the temperature range of the isotropic phase. The sample was tilted with respect to the beam propagation direction by ~20 degrees. The data presented in (b) and (c) were taken simultaneously, at a temperature 20 K above the N-Iso phase transition. Note that the threshold electric field inducing a non-zero light transmission coincides with the field at which the current peak is registered.

As the system approaches the critical end point in the temperature-electric field phase diagram, the step-like increase in the birefringence at the threshold field becomes smaller and the hysteresis width diminishes. Above the critical temperature, $T_c$, the phase transition vanishes and the evolution of the birefringence in the electric field becomes continuous (Fig. 2). The threshold field – temperature dependence is nearly linear, and the critical point is located at $T_c$ nearly 30 K above the zero-field transition temperature and at a critical electric field $E_c \sim 15$ V/μm (Fig. 3). On the critical isotherm, at the electric field corresponding to the inflection point, the retardation is approximately half of that measured in the $N_F$ phase at the same experimental conditions, which shows that the critical nematic order parameter is $S_c \sim 0.45$ (in the $N_F$ phase far from transition temperature, $S \sim 0.9$ has been previously determined [9]).

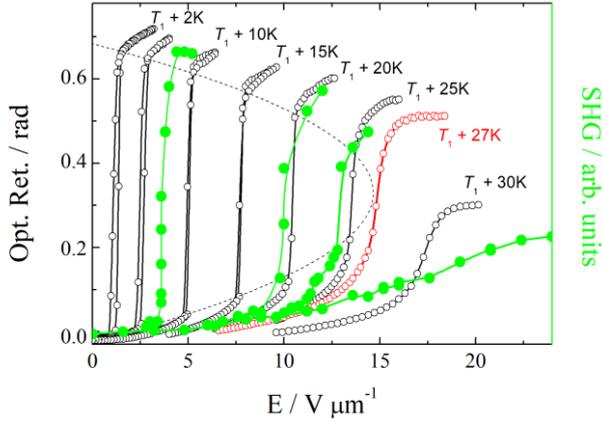

Fig 2. The optical retardation (black, open symbols) and SGH signal intensity (green, solid symbols) measured at several temperatures above $T_1$, the N-Iso phase transition as a function of a dc electric field applied to the sample. The critical isotherm for retardation is marked in red. A dashed curve represents a bimodal line.

To exclude the possibility that the birefringence induced by the electric field is due to the Kerr effect (thus due to a quadratic coupling of the electric field to the dielectric anisotropy), we also studied the second harmonic generation (SHG) to evidence a non-centrosymmetric (polar) structure of the induced phase. In the $N_F$ phase, a strong SHG signal is registered. In a cell with comb-like electrodes in which the electric field and thus polarization in the neighboring areas is antiparallel, the diffraction of the SHG signal is observed at an angle corresponding to double the distance between the electrodes (see Fig. S4 in SM). The SHG signal intensity is only weakly dependent on temperature. In the isotropic liquid phase, the diffraction SHG signal could also be induced by the application of an electric field, but in order to compare the threshold for induction of the SHG signal and for induction of a birefringence, exactly the same thin ITO cells with electrodes on glass surfaces were used in both experiments and the sample was placed at an oblique angle to the incident light. The SHG signal intensity measured as a function of the temperature and applied dc electric field generally follows the same dependence as the induced birefringence (Fig. 2). At each temperature the offset of the SHG signal coincides with the threshold electric field at which the birefringence is induced; as the temperature increases the electric field necessary to induce the step-like growth of the SHG signal intensity increases.

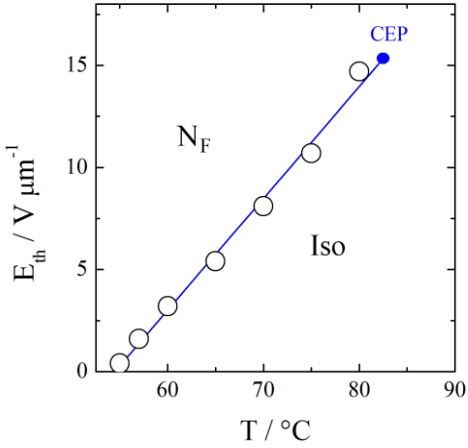

Fig 3. Temperature ($T$) – threshold field ($E_{th}$) phase diagram for the N$_F$ and Iso phases. The critical end point (CEP) is located approximately 30 K above the zero-field transition temperature and at critical electric field ~15 V/micron.

To analyze the Iso-N-N$_F$ phase transitions, we use a phenomenological Landau model, which can describe these phase transitions in a unified way. Characteristics of the nematic phase is the orientation of the long molecular axes along some common director **n** ($\mathbf{n}^2 = 1$). The degree of this orientational order is described by the order parameter $S$, where $0 < S < 1$. A polar nematic phase additionally possesses a polar order of the longitudinal component of dipole moments, which is described by an additional order parameter, spontaneous polarization $P$. We assume that the polarization direction is along the director and thus consider $P$ as a simple scalar order parameter, which, however, is coupled to the nematic order, because ordering of the dipoles also essentially introduces an ordering of long molecular axes. For the materials studied so far, the polar nematic phase exhibits a strong orientational order of molecules (e.g. for RM734, the N$_F$ phase appears when the order parameter S reaches ~0.75 [9]), thus we assume that the polar ordering is not critical and it becomes energetically favored when the nematic order is sufficiently high ($S = S_0$). To analyze the influence of an electric field ($E$), we assume a linear coupling between the polarization and the electric field. Therefore, the free energy ($G$) is expressed as:

$$G = \frac{1}{2}a_S(T-T_0)S^2 + \frac{1}{3}b_S S^3 + \frac{1}{4}c_S S^4 + \frac{1}{2}a_p(S_0-S)p^2 + \frac{1}{4}b_p S^2 p^4 - EP_0 p \qquad (1)$$

where $p = P/P_0$ is a polarization order parameter and $P_0$ is the maximum value of polarization (all longitudinal dipole moments in the same direction), thus $p < 1$. The parameters $a_S$, $b_S$, $c_S$, $a_p$ and $b_p$ are material parameters and are assumed to be temperature ($T$) independent. The Iso-N transition is first order if $b_S < 0$, the rest of the parameters are positive. The term with $b_p > 0$ stabilizes a finite polarization value at $S > S_0$. Chiral terms were omitted in the free energy expresion, eq. (1), because for the studied system they do not to have much influence on the phase behavior, as discussed above. By analyzing the free energy $G$ for stability limits of the phases in the absence of an applied electric field, the model predicts that the isotropic phase becomes unstable on cooling at $T = T_0$, while the nematic phase remains (meta)stable until $T_1 = T_0 + b_s^2/(4a_S c_S)$.

We are interested in the (meta)stability of the polar isotropic phase and polar nematic phase induced by the application of an external electric field at temperatures above $T_1$. For this purpose, we minimize the free energy $G$ over $S$ and $p$ at a given temperature ($T$) and external field $E$. From the conditions $\partial G/\partial S = 0$ and $\partial G/\partial p = 0$ we find a set of coupled equations for the equilibrium values of $p$ and $S$:

$$p = \sqrt{\frac{a_p - \sqrt{a_p^2 - 8b_p S^2 (a_S \Delta T + c_S S^2 + b_S S)}}{2 b_p S}} \qquad (2)$$

and

$$EP_0 = b_p p^3 S^2 + a_p p (S_0 - S). \qquad (3)$$

The dependence of $S$ on $E$ at some $\Delta T$ is best presented by plotting $S(E)$ for a set of $\Delta T$ at a chosen set of parameters, for which we choose $b_S/a_S = -20$ K, $c_S/a_S = 60$ K, $a_p/a_S = 30$ K, $b_p/a_S = 30$ K and $S_0 = 0.5$. For this set of parameters $T_1 - T_0 = 1.67$ K. The parameters were chosen such that the critical temperature is approximately 20 K above $T_1$, the bistability region of the Iso and N phases in the absence of an external field is of the order of 1 K and the width of hysteresis at temperatures above but close to $T_1$ is of the order of $0.1\, E_c$. From Fig. 4 we see that the curves $S(E)$ are S-shape, which implies a discontinuous Iso-$N_F$ transition with a hysteresis loop. The hysteresis loop is defined by stability limits of two minima in the free energy vs $S$. The first-order transition vanishes at a critical temperature $T_c$, where the $S(E)$ curve exhibits a vertical inflection point. Above $T_c$ no transition is detected, only a smooth increase of the orientational order parameter (and $p$) with increasing electric field is observed. By solving a set of equations $\partial E/\partial S = 0$ and $\partial^2 E/\partial S^2 = 0$, we find the critical temperature and field to be $T_c - T_0 = 19.3$ K and $E_c P_0/a_S = 4.64$ K.

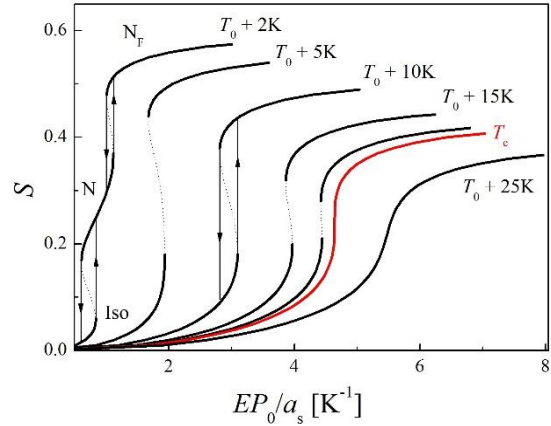

Figure 4. The order parameter $S$ vs. electric field $E$ obtained from eq. (3) at few temperatures above $T_0$. For the chosen set of parameters, the stability limit of the N phase is at $T_1 = T_0 + 1.67$ K and the critical temperature for the Iso-$N_F$ transition at $T_c = T_0 + 19.3$ K. At $T = T_0 + 2$ K the electric field induces a transition from the Iso to the N phase and at a sufficiently high field to the $N_F$ phase. At higher temperatures, a direct Iso-$N_F$ phase transition is induced by the electric field, with a diminishing hysteresis as $T_c$ is approached; the critical isotherm is marked in red. Above $T_c$ a continuous evolution of the order parameter is observed under the electric field.

With a chosen set of parameters a temperature-electric field phase diagram for the N, $N_F$, and Iso phases (Fig. 5) was obtained that can mimic the behavior of the experimental system. In the phase diagram, we included the stability limits for the isotropic, nematic and polar nematic phases vs. electric field and temperature. The lines for the field-induced Iso-$N_F$ and $N_F$-Iso phase transitions, showing the width of hysteresis, join at the critical end point. The dependence of the threshold field $E_{th}$ for the onset of the $N_F$ phase vs. temperature is nearly linear, as observed experimentally. The phase diagram also foresees the induction of the N phase at temperatures close to $T_1$, which was not observed experimentally, most probably due to the fact that the studied material had a very narrow temperature range of bistability of the Iso and N phase at zero field as well as a very narrow temperature window of the stability of the N phase. To observe the induction of the $N_F$ phase from the N phase one would need a system with a larger temperature range of the N phase but still a strong first-order N-$N_F$ phase transition.

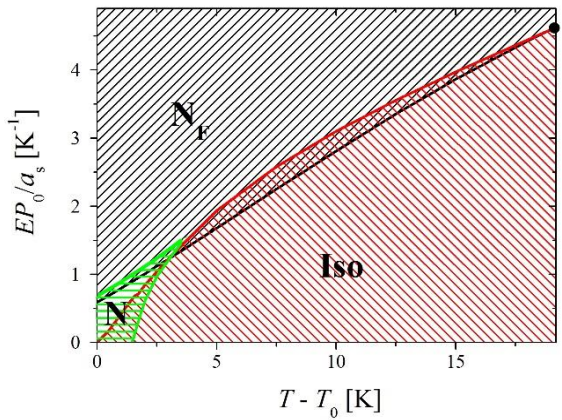

Figure 5. A theoretical $E$ vs. $T$ phase diagram showing the stability limits of the Iso (red area), N (green area), and $N_F$ (black area) phases. In the areas marked with two colors, coexistence of two phases is possible due to the presence of two minima in the free energy $G$ vs. $S$ dependence under electric field. The black dot shows the critical end point for the induced Iso-$N_F$ phase transition.

In conclusion, we have investigated the influence of an electric field on the transition between the isotropic and ferroelectric nematic phase at temperatures above the stability of the nematic phase at zero field. The phase transition is first-order for the system studied. The overall experimental data show that the behavior of the ferroelectric nematic phase is very similar to that of a liquid near the liquid-gas critical point [13] above which a continuous evolution is observed between these two states of matter, a polar order and electric field strength corresponding to density and pressure, respectively. It should also be pointed out that the observed behavior is similar to the behavior of systems with the SmA-SmC* [14] or SmA-Iso [15] phase transitions, in materials in which these transitions are first-order in nature. However, in smectic systems, the critical end point was located less than 1 K above the zero-field transition temperature, while in the $N_F$ system, it is located approximately 30 K above. This makes such systems interesting from the application point of view because a strong birefringence can be induced in a broad temperature range in an optically isotropic phase.

## Acknowledgment

The research was supported by the National Science Centre (Poland) under the grant no. 2021/43/B/ST5/00240. NV and MC acknowledge a support of the Slovenian Research Agency (ARRS), through the research core funding program No. P1-0055.

# Supplementary Material

**Experimental:**

*Electrooptical Measurements:* The induced optical retardation was measured with a setup consisting of a photoelastic modulator (PEM-90, Hinds), halogen lamp (Hamamatsu LC8) equipped with a narrow bandpass filter (532 nm) as a light source, and a photodiode (FLC Electronics PIN-20). The measured intensity of the transmitted light was de-convoluted with a lock-in amplifier (EG&G 7265) into 1f and 2f components to yield a retardation induced by the sample. Glass cells with a thickness of 5-10 μm and transparent ITO electrodes were used. Measurements were performed as a function of an applied electric bias field which, above the threshold value, induced a homeotropic alignment of the director in the induced $N_F$ phase. The cells were placed at an oblique angle with respect to the light propagation direction.

The current peaks due to a spontaneous electric polarization reversal were recorded upon applying a triangular voltage with a frequency of 4 Hz by monitoring a voltage drop on a resistor connected in series with the cell. Simultaneously, the light transmission through the tilted cell placed between crossed polarizers was recorded with a photodiode. For chosen temperatures, the current peaks were also recorded for higher frequencies up to 50Hz.

*SHG measurements:* The SHG response was investigated using a setup based on a solid-state laser EKSPLA NL202. 9 ns laser pulses at a 10 Hz repetition rate and ~2 mJ pulse energy at λ=1064 nm were applied. The pulse energy was adjusted for each sample to avoid its decomposition. The infra-red beam was incident onto a LC homogenous cell of thickness 1.7-20 μm. An IR pass filter was placed at the entrance to the sample and a green pass filter at the exit of the sample. The emitted SHG radiation was detected using a photon counting head (Hamamatsu H7421) with a power supply unit (C8137). The signal intensity was estimated by a custom-written Python script reading the oscilloscope output signal (Agilent Technologies DSO6034A).

**Supplementary Information:**

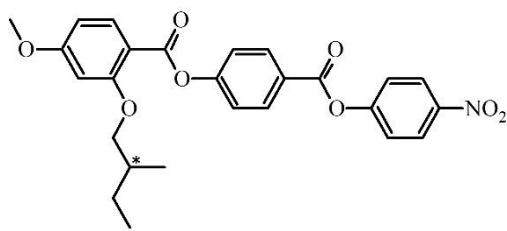

**Figure S1.** The chemical formula of the studied material I-4*. Phase transition temperatures (in °C), determined by DSC are, for optically pure compound and: m.p. 140.0, Iso-N 56.1 N-$N_F$ 54.8 and for racemic mixture: m.p. 140.8, Iso-N 56.0 N-$N_F$ 54.7.

Supplementary Material

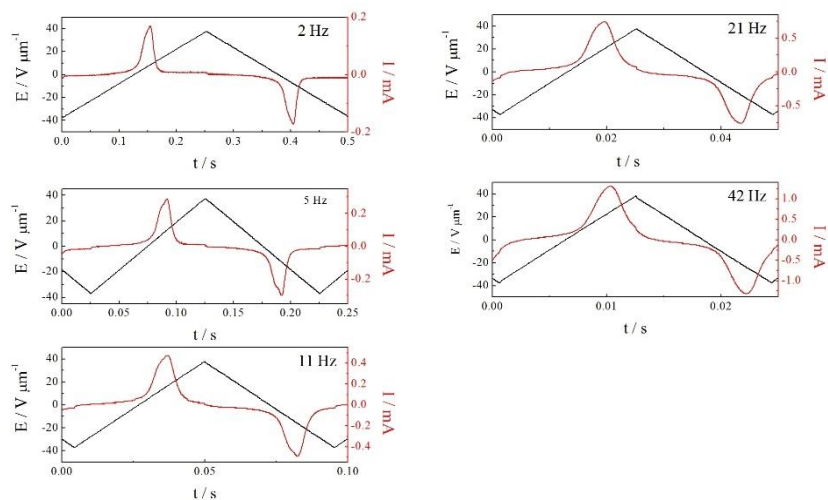

**Figure S2**. Current peaks due to the electric polarization reversal recorded in the $N_F$ phase under an application of a triangular wave voltage of various frequencies (material I-4*, filled into 5-μm-thick cell without polymer layers, T=45 °C). The value of the electric polarization determined by integration of the current peak was 4.2 μC cm$^{-2}$, independent of the frequency of the applied electric field, which proves that a possible contamination of the sample with ionic impurities is low, and does not contribute to the recorded current peaks.

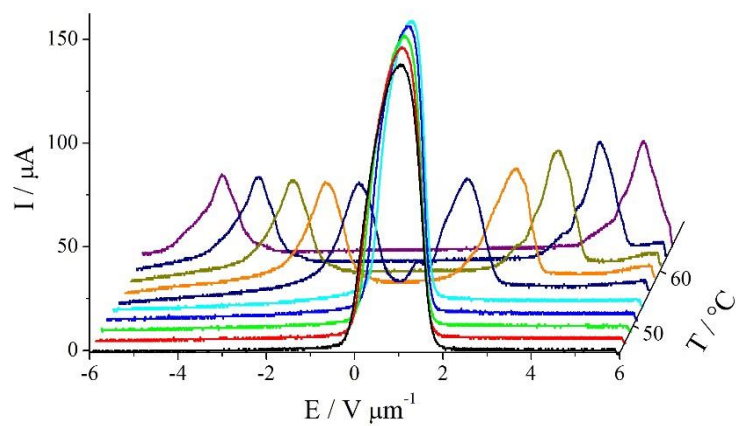

**Figure S3**. Current peaks recorded at various temperatures in the $N_F$ and Iso phases under an application of a triangular wave voltage to the cell (half of the cycle is presented). In the $N_F$ phase, a single current peak corresponds to a reversal of the polarization vector direction, while in the Iso phase, two separate peaks are related to field-induced $N_F$-Iso and Iso-$N_F$ transitions.

Supplementary Material

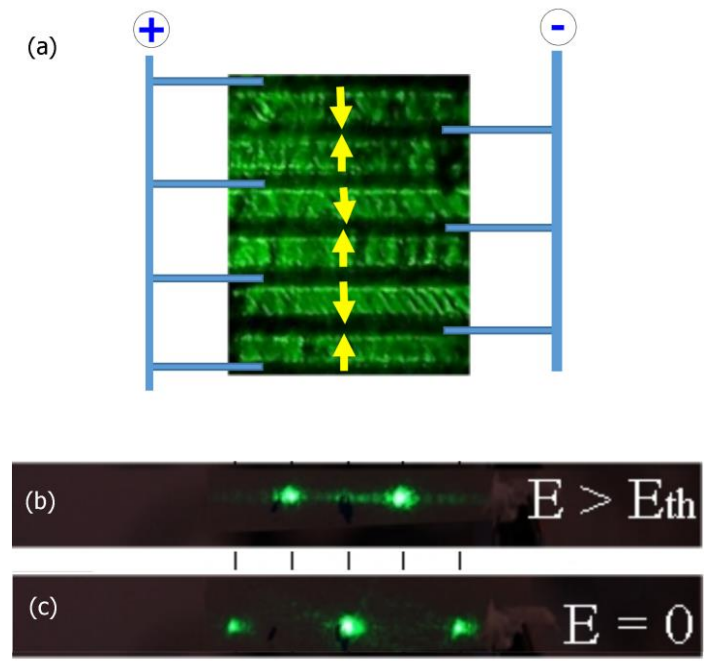

**Figure S4.** (a) An SHG signal recorded in a cell with comb-like electrodes upon an application of an electric field ($E$); the resulting structure of polarization vectors is presented by yellow arrows. (b) Due to an alternating polarization direction in the neighboring inter-electrode areas (slits), the SHG rays emitted from those areas are phase shifted by $\pi$, thus the resulting diffraction pattern corresponds to a grating with a periodicity equal to twice the distance between the electrodes. For the same reason, the zero-order diffraction signal is missing, because the SHG light is emitted from the neighboring slits with opposite phases and a complete destructive interference takes place. (c) For comparison, the SHG diffraction pattern for the sample at zero field ($E = 0$) is recorded – under such a condition the positive interference is observed for a periodicity equal to the distance between electrodes, due to the random orientation of ferroelectric domains in areas between the electrodes, which makes them equivalent.